\documentclass[aps,prl,twocolumn,preprintnumbers,showpacs,onlinecite]{revtex4}
\usepackage{amssymb}
\usepackage{bm}
\usepackage{amsmath}
\usepackage[dvips]{graphicx}
\usepackage{color}

\begin{document}

\title{Magnetic quantum oscillations and multiple holon pockets in underdoped 
YBa$_2$Cu$_3$O$_{6+y}$}

\author{Wei Chen$^{1}$, Oleg P. Sushkov$^{1}$, and Takami Tohyama$^{2}$}

\affiliation{$^{1}$School of Physics, University of New South Wales, Sydney 2052, Australia \\
$^{2}$Yukawa Institute for Theoretical Physics, Kyoto University, Kyoto 606-8502, Japan}

\date{\today}

\begin{abstract}
We show that lightly doped YBa$_2$Cu$_3$O$_{6+y}$ 
has multiple holon pockets with different areas,
 which lead to multiple frequencies of magnetic quantum oscillations. 
Using neutron scattering data on incommensurate spin ordering
we determine these areas, which yields frequencies in good 
agreement with experiments. Divergence of the effective mass observed in 
magnetic quantum oscillations indicates a quantum phase transition at the 
oxygen content $y\approx 0.48$. We argue that the transition is the onset 
of quasistatic incommensurate  magnetic order
predicted by theory and observed  in neutron scattering.
\end{abstract}

\date{\today}


\maketitle
Topology of the Fermi surface is one of the central problems
in  the physics of cuprate superconductors.
An undoped cuprate is antiferromagnetic Mott insulator.
At a sufficiently small doping the system can be  described by the extended 
two-dimensional $t$-$J$ model~\cite{PWA,ZR} that
predicts small holon pockets located at the ``nodal points''.
We use the term ``holon'' to stress that the quasiparticle does not carry 
usual spin, with more details to be explained later. 
On the other hand there is no doubt that at a sufficiently large 
doping the system 
behaves like a normal metal with a large Fermi surface and normal 
quasiparticles which carry simultaneously electric charge and spin $S$=1/2. 
Description of the transition between these two regimes is an
open theoretical problem.

Angle resolved photoemission spectroscopy
(ARPES) studies 
indicate a large Fermi surface in overdoped regime and formation of 
Fermi arcs in the underdoped regime~\cite{Norman98}.
On the other hand, recent magnetic quantum oscillation (MQO) 
data~\cite{Doiron-Leyraud07,LeBoeuf07,
Yelland08,Jaudet08,Sebastian08, 
Audouard09,Sebastian09,Sebastian09_2,Singleton09}
taken in YBa$_2$Cu$_3$O$_{6+y}$(YBCO)
indicate small Fermi pockets at $y < 0.66$ that corresponds to 
doping $x < 0.125$.
Throughout the paper we denote doping by $x$ and
use results of Ref.~\onlinecite{Liang} to relate  the oxygen content $y$
with the doping level $x$.
Very recent ARPES studies~\cite{Meng09,He09} also give some indications 
 of small Fermi pockets.
A theoretical interpretation of ARPES in 
terms of holons is a fairly involved issue. 
In a photoemission measurement partially separated spin and charge have to
recombine to form a physical electron. 
So far the recombination amplitude has only been calculated for the parent
Mott insulator~\cite{sushkov97}.
In contrast, MQO is sensitive mainly to the electric charge,
hence interpretation of MQO in terms of holons is 
straightforward. With tilted magnetic field MQO can also probe spin of the 
quasiparticle. A very recent measurement of this kind~\cite{Sebastian09} 
indicates suppression of spin and hence supports the holon picture.

The sign of the Hall coefficient measured 
in Ref.~\onlinecite{LeBoeuf07} corresponds to electron 
pockets instead of hole ones.
However,  the field applied in the experiments 
is smaller than the critical field $H_{c2}$, hence the contribution from the 
vortex liquid has to be taken into account, which may alter sign of  the
Hall coefficient~\cite{Galffy88,Khom95}. 
In the present work we consider only hole pockets.

In this paper we address the following issues related to MQO: 
(1)Typically more than one MQO frequency is observed. What is the origin for 
the multiple frequencies? 
(2)The main MQO frequency corresponds to the area of the pocket about 1.8\% 
of the Brillouin zone. How to reconcile this very small area with 
the  Luttinger sum rule? 
(3)Divergence of the effective mass has
been observed~\cite{Sebastian09_2} at doping about 9\%. 
What is the physical origin for this quantum
critical point (QCP)?

We will relate MQO with recent neutron scattering 
observations~\cite{Stock04,Hinkov07,Hinkov08,Haug09,Haug09a}.
These data demonstrate incommensurate spin ordering
pined to the tetragonal $a^*$ direction. Depending on doping the ordering 
can be static or dynamic. The QCP separating regions of dynamic
and static ordering is located at  $x_\mathrm{QCP}\approx 0.09$. 
The spin-wave pseudogap $\Delta_\mathrm{sw}$ is opened at $x > x_\mathrm{QCP}$. 
The data~\cite{Stock04} on YBa$_2$Cu$_3$O$_{6.5}$ ($x\approx 0.1$) 
show $\Delta_\mathrm{sw} \sim 10$~meV,
while in YBa$_2$Cu$_3$O$_{6.6}$ ($x\approx 0.12$)
$\Delta_\mathrm{sw} \sim 20$~meV~\cite{Hinkov07}.
On the other hand the 
quasistatic scattering has been 
observed at $x < x_\mathrm{QCP}$ where $\Delta_\mathrm{sw}=0$.
In YBa$_2$Cu$_3$O$_{6.45}$ ($x\approx 0.085$)
the signal is very weak~\cite{Hinkov08,Haug09}, and it is much 
bigger~\cite{Haug09a} in YBa$_2$Cu$_3$O$_{6.35}$ ($x\approx 0.065$).

MQO in a magnetic field $B$ is described as 
$ \cos(2\pi\frac{F}{B}+\phi)$.
The period $F$ and the area enclosed by a trajectory in the momentum space 
$A_k$ are related as~\cite{Ziman}
\begin{eqnarray}
\label{FA}
F=\frac{c\hbar}{2\pi e}A_k \ .
\end{eqnarray}
The typical measured value of $F$ in YBCO is about 500~T,
which gives a ratio of $A_k$ to the total area of the Brillouin zone
$A_\mathrm{BZ}=(2\pi/a)^2$ ($a=3.81$\thinspace \AA $\,$ is the lattice spacing)
to be $A_k/A_\mathrm{BZ}\approx 0.0176$.

Our analysis of MQO is based on the spin spiral theory of a lightly 
doped Mott insulator. While the
idea of the spin spiral has been suggested quite some time ago
\cite{shraiman88,IF,CM,sushkov04}, 
the consistent theory with account of quantum fluctuations
has been developed only recently~\cite{Milstein08}. 
Generalization of the spiral theory to bilayer YBCO
has been discussed in  Ref.~\onlinecite{Sushkov09}.
The theory is based on expansion in powers of doping $x$,
so it is parametrically justified at $x \ll 1$.
 The theory is formulated in terms of  
the bosonic ${\vec n}$-field ($n^2=1$) 
that describes the staggered component of the  copper spins,
and in terms of  fermionic holons $\psi$.
The holon field $\psi$ carries electric charge and it has a pseudospin that originates from two sublattices.
Minima of the holon dispersion are at the nodal points
$\mathbf{q}_{0}=(\pm \pi /2,\pm \pi /2)$,
so there are holons of two types $\psi_{\alpha}$, $\alpha =1,2$, 
corresponding to two nodal directions.
The dispersion in a pocket is somewhat
anisotropic, but for simplicity we use here the isotropic approximation,
$\epsilon \left( \mathbf{p}\right) \approx \beta \mathbf{p}^{2}/2$,
where ${\bf p}={\bf q}-\mathbf{q}_{0}$.
Lattice spacing of the square lattice is set to be equal to unity 
$a=3.81$\thinspace \AA $\,\rightarrow $\thinspace 1. 
The effective Lagrangian for a single layer system
reads~\cite{Milstein08}
\begin{eqnarray}
\label{eq:LL}
{\cal L}&=&\frac{\chi_{\perp}}{2}{\dot {\vec n}}^{2}-\frac{\rho_{s}}{2}(\nabla {\vec n})^{2}+\sum_{\alpha}\{\frac{i}{2}[\psi_{\alpha}^{\dag}{\cal D}_{t}\psi_{\alpha}-({\cal D}_{t}\psi_{\alpha})^{\dag}\psi_{\alpha}]
\nonumber \\
&-&\psi_{\alpha}^{\dag}\epsilon({\cal P})\psi_{\alpha}+\sqrt{2}g(\psi_{\alpha}^{\dag}{\vec \sigma}\psi_{\alpha})\cdot[{\vec n}\times({\bf e_{\alpha}\cdot\nabla}){\vec n}]\}\ ,\nonumber\\
{\cal P}&=&-i{\bf \nabla}+\frac{1}{2}\vec{\sigma}\cdot[\vec{n}\times {\bf \nabla}\vec{n}]\;,\nonumber\\
{\cal D}_{t}&=&\partial_{t}+\frac{i}{2}\vec{\sigma}\cdot[\vec{n}\times\dot{\vec{n}}]\;.
\end{eqnarray}
First two terms in the Lagrangian represent the usual nonlinear 
$\sigma$ model, with magnetic susceptibility $\chi_{\perp}\approx 0.066 J$ 
and spin stiffness $\rho_s \approx 0.175J$,
where $J\approx 130$~meV is the antiferromagnetic exchange in 
the parent Mott insulator.
The extended $t$-$J$ model predicts the following
values of the coupling constant and the inverse
effective mass, $g \approx 1.0J$, $\beta\approx 2.5J$
($m^*=1.6m_e$).
Below we will not use this value of $\beta$,
instead we will treat $\beta$ as a fitting parameter.
The pseudospin operator is $\frac{1}{2}{\vec \sigma}$,  and 
${\bf e}_{\alpha}=(1/\sqrt{2},\pm 1/\sqrt{2})$ is a unit  vector orthogonal to the face of the magnetic Brillouin zone (MBZ) where the holon is located. 
Note that usage of the MBZ notations  does not imply that there is a long range
magnetic order. This is just a convenient way to avoid double counting
of degrees of freedom, and the pseudospin accounts for doubling of 
the area of MBZ. Due to the MBZ notations one should consider two 
full pockets located at $\mathbf{q}_{0}=(+ \pi /2,- \pi /2)$ and 
$\mathbf{q}_{0}=(+ \pi /2,+ \pi /2)$, hence the index $\alpha$ takes two 
values. We stress that it does not matter if the ground state 
expectation
value of the n-field is nonzero, $\langle {\vec n}\rangle\ne 0$
(magnetic ordering), 
or zero, $\langle {\vec n}\rangle= 0$.
The only condition for validity of  Eq.(\ref{eq:LL}) is that  dynamic 
fluctuations of the ${\vec n}$-field are sufficiently slow.
Typical energy scale of the ${\vec n}$-field quantum fluctuations is 
$E_\mathrm{cross}\propto x^{3/2}$ (position of the neck of the
``hour glass'' spin wave dispersion),
as discussed in Ref.~\onlinecite{Milstein08}, and it must be small compared 
to the holon Fermi energy $\epsilon_F \propto x$. The inequality 
$E_\mathrm{cross} \ll \epsilon_F$ is valid up to optimal doping,
$x \approx 0.15$, below which Eq.(\ref{eq:LL}) is parametrically 
justified.

To account for the interaction with magnetic field, we make the 
following modifications in Eq.(\ref{eq:LL})~\cite{Milstein08}: 
A magnetic vector potential ${\bf A}$ is included in the long derivative 
${\cal P} \to {\cal P}-\frac{e}{c}{\bf A}$, kinetic energy of $n-$field is 
modified ${\dot {\vec n}}^{2} 
\to \left(\dot {\vec n}-[{\vec n}\times{\vec B}]\right)^{2}$,
and an extra term
$\delta{\cal L}_B=\frac{1}{2}({\vec B}\cdot{\vec n})
\psi^{\dag}_{\alpha}({\vec \sigma}\cdot{\vec n})\psi_{\alpha}$ is included.
Here we include the Bohr magneton in definition of the magnetic field,
$2\mu_\mathrm{B}B \to B$. The precise meaning of spin-charge separation is clear under 
this context: In a normal Fermi liquid the spin interaction 
is $\delta{\cal L}_{B}^\mathrm{NFL}=\frac{1}{2}
\psi^{\dag}_{\alpha}{\vec B}\cdot{\vec \sigma}\psi_{\alpha}$.
In contrast, for a spin-spiral state
the $g$-term in Eq.(\ref{eq:LL})
enforces ${\vec n} \perp {\vec \sigma}$, hence the expectation value
of $\delta{\cal L}_{B}$ goes to zero, 
$\langle ({\vec B}\cdot{\vec n})({\vec \sigma}\cdot{\vec n})\rangle =0$. 
Thus the pseudospin does not
interact with magnetic field in the first order in $B$ and this is the
meaning of the spinless nature of the holon.

\begin{figure}[h]
\begin{center}
\includegraphics[clip=true,width=0.99\columnwidth]{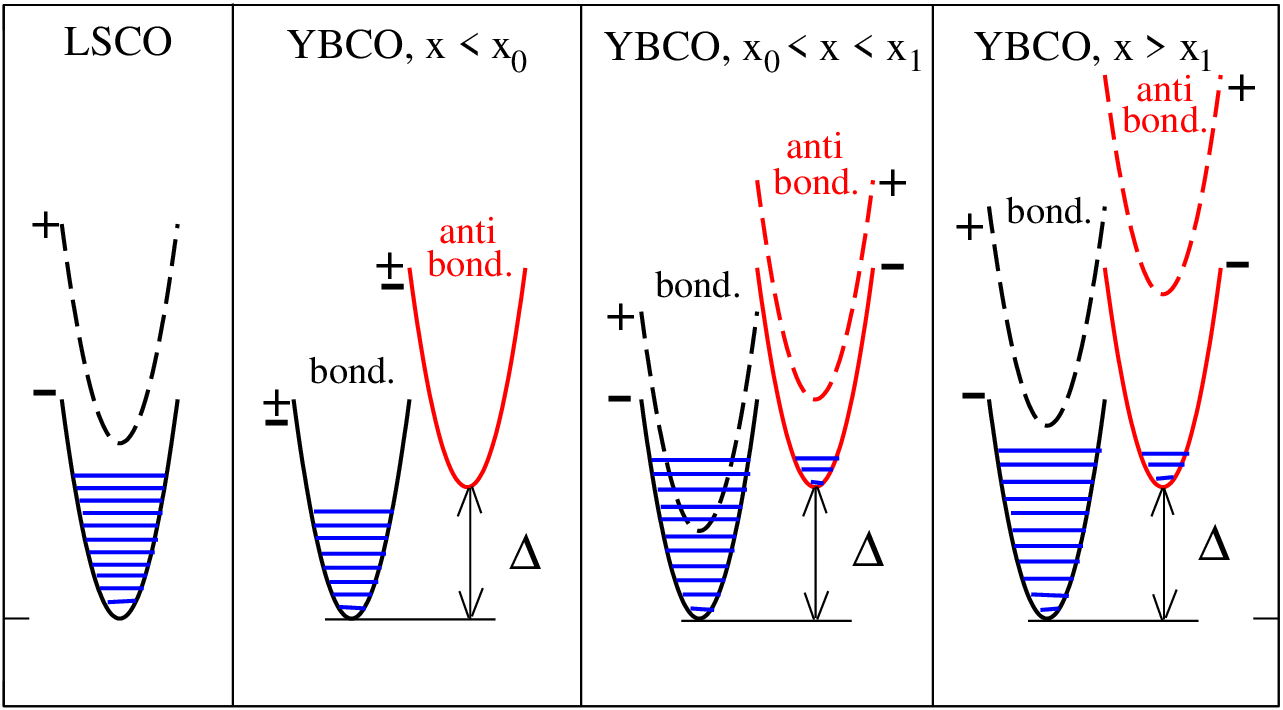}
\caption{(color online) Filling of holon bands.
Left panel:  single layer La$_{2-x}$Sr$_x$CuO$_4$ (LSCO).
Three right panels: double layer YBCO in three different regimes.
The solid and the dashed line correspond to different pseudospin
projections, the splitting is $\pm gQ$. In the YBCO panels left/right 
parts show bonding/antibonding bands, the splitting is $\pm\Delta/2$.
}
\label{disp11}
\end{center}
\vspace{-20pt}
\end{figure}

In the effective action (\ref{eq:LL}) the $\rho_s$-term, the 
 $\epsilon({\cal P})$-term and the $g$-term are important in
semiclassical approximation. All other terms contain time derivatives,
which are important only for quantum fluctuations and we disregard 
them for now.
For the single layer case, we choose the plane of the coplanar spin spiral 
to be the
$xy$-plane in the spin space,  
${\vec n}=(\cos{\bm Q}\cdot{\bm r},\sin{\bm Q}\cdot{\bm r},0)$.
Due to the $g$-term in Eq.(\ref{eq:LL}) the psedospin of a
holon is quantized along z-axis. Energy of the holon 
is $\epsilon({\bf p})=\pm gQ+\beta {\bf p}^2/2$,
where $\pm gQ$ is the  pseudospin  splitting.
The splitting between two pseudospin branches is shown in the left panel
of Fig.\ref{disp11}.
Minimization of the semiclassical energy shows that ${\bm Q}$ is directed along
the CuO bond [${\bm Q}\propto (1,0)$ or ${\bm Q}\propto (0,1)$] and has value
$Q=\frac{g}{\rho_s}x$.
The upper pseudospin branch of the holon dispersion is always empty,
as shown in the left panel, Fig.\ref{disp11}, hence the area of the holon 
pocket is $\frac{A_k}{A_\mathrm{BZ}}=\frac{1}{2}x$.
Quantum fluctuations reduce the static value of spin, 
$|\langle {\vec n}\rangle| < 1$, moreover, at $x > x_\mathrm{QCP}\approx 0.1$
the expectation value of ${\vec n}$ vanishes,
$\langle {\vec n}\rangle =0$, and the
spiral becomes fully dynamic~\cite{Milstein08}.
However, this does not influence the value of the incommensurate vector
${\bm Q}$ because the fluctuations are slow.
The point is that the semiclassical analysis is based on relatively
short distance and time, with corresponding typical momenta and 
energies
$Q < q_\mathrm{semi} < p_F \propto \sqrt{x}$,
$\omega_\mathrm{semi} \sim \epsilon_F \propto x$.
Quantum fluctuations come from smaller momentum/energy scales
 $q_\mathrm{quant} < Q$, 
$\omega_\mathrm{quant} \propto x^{3/2}$, as discussed in Ref.~\onlinecite{Milstein08}.

In the double layer case, one has to include the interlayer hopping in 
the effective action (\ref{eq:LL}). As a result, holon 
wave function $\psi$ attains bonding/antibonding
index  with respect to the interlayer hybridization~\cite{Sushkov09}.
The holon energy at {\it each of the two
nodal points} reads
$\epsilon({\bf p})=\beta{\bf p}^{2}/2\pm gQ \pm\Delta/2$,
where $\pm gQ$ is the  pseudospin  splitting  
and $\pm \Delta/2$ is the antibonding(a)/bonding(b) splitting.
Effectively there are {\it four} different bands per nodal direction
($b-$, $b+$, $a-$, $a+$),
altogether {\it eight} bands. The filling configuration of these bands 
is determined by minimizing the semiclassical energy, which yields three 
different doping regimes~\cite{Sushkov09}
\begin{eqnarray}
\label{Q2}
&& 1)\ \ x < x_0\ ,\ \ \ \ \ \ \ \ Q=0  \nonumber\\
&& 2)\ \ x_0 < x < x_1\ ,  \ 
Q=\frac{g}{\rho_s}\frac{x-\Delta/(\pi\beta)}{3-2\lambda} \nonumber\\
&& 3) \ \ x > x_1\ , \ \ \ \ \ \ \ \ Q=  g x /\rho_s \ ,
\end{eqnarray}
where $\lambda=2g^2/(\pi\beta\rho_s)$.
The points $x_0$ and $x_1$ are Lifshitz points. Filling configuration of these 
three regimes are shown in Fig.\ref{disp11}.
Areas of the filled holon pockets are
\begin{eqnarray}
\label{AA}
 1)\ \ \ A_{b-}/A_\mathrm{BZ}&=&A_{b+}/A_\mathrm{BZ}=x/2  
\nonumber\\
2)\ \ \
A_{b-}/A_\mathrm{BZ}&=&
x/3+gQ/(3\pi\beta)+\Delta/(6\pi\beta)\;
\nonumber \\
A_{b+}/A_\mathrm{BZ}&=&
x/3-2gQ/(3\pi\beta)+\Delta/6\pi\beta)\;
\nonumber \\
A_{a-}/A_\mathrm{BZ}&=&
x/3+gQ/(3\pi\beta)-\Delta/(3\pi\beta)\;
 \nonumber\\
3) \ \ \ A_{b-}/A_\mathrm{BZ}&=&
x/2+\Delta/(4\pi\beta)\;
\nonumber \\
A_{a-}/A_\mathrm{BZ}&=&
x/2-\Delta/(4\pi\beta)\;.
\end{eqnarray}
Note that the total occupied area is $2x$ because there are two layers.
From (\ref{AA}) we conclude that there is one MQO frequency in the first 
regime, three frequencies in
the second regime, and two frequencies in
the third regime. Let us  consider two scenarios for the doping dependence 
of the hybridization splitting: 
(A) a constant gap $\Delta=\Delta_0$,
(B) $\Delta=\Delta_0(1+\nu x)$, since the splitting 
$\Delta$ is due to tunneling via interlayer oxygen chains~\cite{andersen95}, 
a linear dependence on doping is possible.

\begin{figure}[th]
\begin{center}
\includegraphics[clip=true,width=0.6\columnwidth]{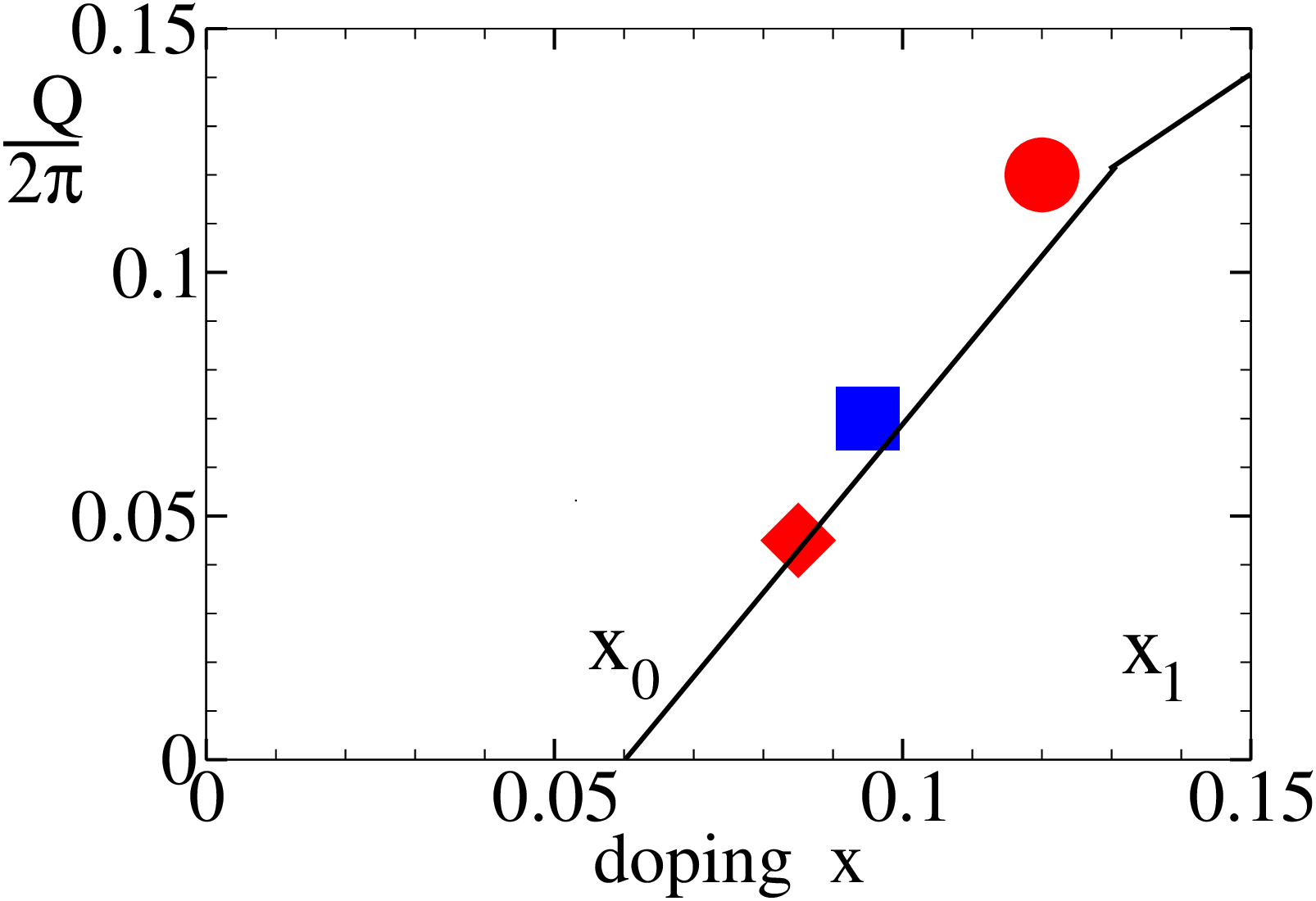}
\vspace{-10pt}
\caption{(color online) 
Incommensurate wave vector versus doping.
The blue square~\cite{Stock04},
the red circle~\cite{Hinkov07},
and the red diamond~\cite{Hinkov08}
show neutron scattering data.
The solid line shows fit of the data using Eqs.(\ref{Q2}).\\
}
\label{QF}
\end{center}
\vspace{-35pt}
\end{figure}

Values of the incommensurate wave vector $Q$ determined by neutron 
scattering~\cite{Stock04,Hinkov07,Hinkov08} are presented in
Fig.\ref{QF}.
Comparing Fig.\ref{QF} with Eqs.(\ref{Q2}) one finds $x_0\approx 0.06$,
$x_1\approx 0.13$.
 Using Eqs.(\ref{Q2}), we can determine
parameters of the model. We fix $g=J$ and $\rho_s=0.175J$, as they are 
predicted by the extended $t$-$J$ model.
Within the scenario (A) the fit gives,
$\lambda=1.23$,  $\beta=2.95J$ ($m^*=1.35m_e$), $\Delta_0=0.556J$.
Within the scenario (B), the data from Fig.\ref{QF} are not sufficient to 
determine the additional parameter $\nu$.
However, we can assume that value of $\lambda$ in YBCO and LSCO is the same,
and use the value $\lambda=1.31$ obtained in Ref.~\cite{Milstein08}
from fitting the data for LSCO.
This gives $\beta=2.78J$ ($m^*=1.43m_e$), $\Delta_0=0.37J$,
$\nu=6.9$ within the scenario (B).
Note that values of  $\beta$ in both fits are close to that predicted by 
the extended $t$-$J$ model, as demonstrated in the paragraph after Eq.(\ref{eq:LL}), and the corresponding effective masses are close to that measured in MQO away from the QCP~\cite{Sebastian09_2,Singleton09}, $m^*=1.6\pm 0.1m_e$.
\begin{figure}[th]
\begin{center}
\includegraphics[clip=true,width=0.9\columnwidth]{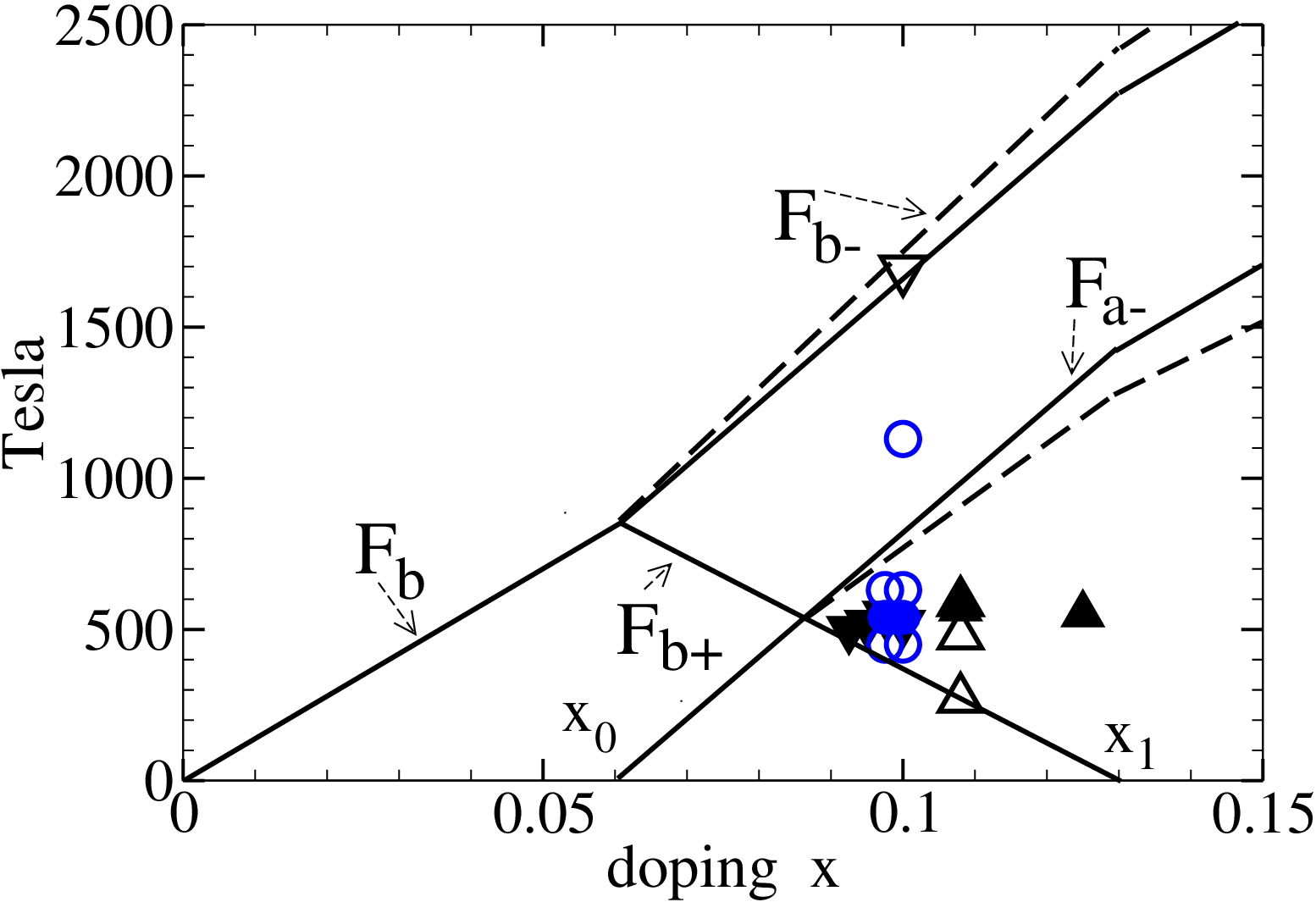}
\vspace{-10pt}
\caption{(color online) 
Frequencies of MQO in Tesla  versus doping.
Solid lines show the theoretical prediction within the
scenario (A), dashed lines show the prediction within
the scenario (B). Predictions (A) and (B) coincide in the lower
part of the figure.
Experimental data are shown by symbols:
Ref.~\cite{Audouard09} - blue circles, 
Ref.~\cite{Sebastian09_2} - black triangles down, 
Ref.~\cite{Singleton09} - black triangles up.
Full symbols  correspond to maximum intensity lines.
}
\label{QF1}
\end{center}
\vspace{-20pt}
\end{figure}

Frequencies of MQO are determined by these parameters via 
Eqs.(\ref{AA}) and (\ref{FA}). 
The results are presented in Fig.\ref{QF1}. 
Difference between the scenario (A) (solid lines) and the scenario (B) 
(dashed lines) is fairly small.
There is quite a reasonable agreement with experimental data, especially having
in mind that the theory has no fitting parameters related to MQO.
Even a high frequency point $\approx 1690T$ is  reproduced.
Notice that MQO always contain 
higher harmonics.
Most likely the blue circle at $\approx 1100T$ is such a 
harmonic~\cite{Sebastian09_2}.
Filling of each pocket is only a fraction of the total doping, 
this explains why the filling extracted from one single frequency seems to 
violate the Luttinger sum rule. The main frequency at $x\approx 0.09$ 
comes from the $b+$ pocket, which has Fermi energy $\approx 40$~meV, and the 
number of filled Landau levels at 
$B=60$Tesla is $N=\epsilon_F/\hbar\omega_c\approx 6$, $\omega_c$ being the synchrotron frequency. 
Amplitude of MQO is proportional to
$\exp\left\{-\pi \Gamma/\hbar\omega_c\right\} $, 
Ref.~\cite{Wasserman96}. Assuming that the exponent is $\sim 1$, and
relating the impurity broadening to the mean free path  and to the Fermi 
velocity, $\Gamma =\hbar v_F/l$, we estimate the holon mean free path 
$l \sim 70a\sim 270$~\AA.

A divergence of the effective mass at approaching $y=0.49$
from higher doping has been reported~\cite{Sebastian09_2}.
The effective mass has been extracted via temperature dependence of the 
MQO signal. The divergence was interpreted~\cite{Sebastian09_2} as a QCP 
due to a metal-insulator transition.
Here we suggest an alternative explanation for this QCP, based on the
current theory and on the evidences from neutron scattering. 
Besides the Lifshitz points $x_0$ and $x_1$, the spiral theory also predicts 
a QCP at $x_\mathrm{QCP}\approx 0.1$ which separates regions of static and dynamic 
spin spiral. The neutron scattering  
data~\cite{Hinkov07,Hinkov08,Haug09,Stock04} 
demonstrate this QCP in YBCO at doping level 
$x_\mathrm{QCP}\approx 0.09$, see discussion in the introduction.  
As one should expect, the precise position of 
this QCP depends on the applied magnetic field.
The quasistatic neutron
scattering in YBa$_2$Cu$_3$O$_{6.45}$ 
is enhanced in the field~\cite{Haug09}, which indicates that magnetic field 
shifts the QCP towards higher doping.
Most likely in the field 60-80 Tesla used in MQO, the QCP is located
between YBa$_2$Cu$_3$O$_{6.47}$ and YBa$_2$Cu$_3$O$_{6.49}$.
The point is that the field is hardly sufficient to close the
spin-wave pseudogap $\Delta_\mathrm{sw}\sim 10$~meV in 
YBa$_2$Cu$_3$O$_{6.5}$~\cite{Stock04}.
The QCP is driven by the long-wave-length quantum 
fluctuations, while holon pockets are formed at a shorter semiclassical scale.
Therefore MQO frequencies are not sensitive to the magnetic QCP.

Effective mass in a quantum field theory always depends on the
momentum transfer $q$, $m^* \to m^*_q$.
The effective mass extracted from fitting neutron scattering data 
is relevant to the semiclassical
scale $q\sim p_F \propto \sqrt{x}$.  This mass does not
``see'' the magnetic QCP.
However, the amplitude of the MQO signal is formed at the length scale
$l \sim 70a\sim 270$~\AA,
which corresponds to a very small $q$ and the corresponding  effective mass 
is sensitive to the magnetic QCP.
It is known~\cite{Vojta00} that the quasiparticle residue $Z$
vanishes at a magnetic QCP. Since $m^* \propto 1/Z$, this implies that
the effective mass is diverging.
Alternatively one can say that the holon scattering rate from critical
magnons is diverging.
This gives a natural explanation of the divergence observed in 
Ref.~\cite{Sebastian09_2}, and also explains why position of the QCP 
observed in neutron scattering coincides with that observed in MQO:
they are the same QCP.

In summary, we explained the MQO frequencies based on the 
spin-spiral theory of lightly doped Mott insulator.
Fit of the incommensurate neutron scattering data shown in Fig.\ref{QF} 
determines free parameters of the theory, and allows us
to calculate frequencies of MQO as  functions
of doping. In the doping region $0.06 < x < 0.13$ the theory
predicts three different frequencies, which yields a fairly good agreement 
with current experimental data, Fig.\ref{QF1},
 and the Luttinger sum rule is reconciled.
We also argue that the quantum critical point observed in magnetic
quantum oscillations (divergence of the effective mass) is the same
quantum critical point that is observed in neutron scattering,
which signatures the onset of the static incommensurate magnetic order.

We thank D. Haug, V. Hinkov and B. Keimer for important
discussions and for communicating unpublished data.
We  acknowledge useful comments by S. Borisenko and C. Zhang.

\end{document}